\documentclass[11pt,a4paper]{article}

\usepackage[margin=1in]{geometry}

\usepackage{xcolor}

\usepackage{amsmath}
\usepackage{amssymb}
\usepackage{amsfonts}
\usepackage{mathtools}

\usepackage{graphicx}
\usepackage{float}
\usepackage{subcaption}
\usepackage{pgfplots}
\pgfplotsset{compat=1.18}

\usepackage{booktabs}
\usepackage{multirow}

\usepackage[
    backend=biber,
    style=phys,
    articletitle=true,
    biblabel=brackets
]{biblatex}

\usepackage{hyperref}
\hypersetup{
    colorlinks=true,
    linkcolor=blue,
    citecolor=blue,
    urlcolor=blue
}

\title{\textbf{Analyzing $\mu$ Parameter Range and the Duration of Reheating in the Polynomial $\alpha$-attractor Models using CMB Observations}}

\author{%
    \textbf{Ajay Atwal}$^{1}$ \footnote{d23017@students.iitmandi.ac.in, ajayatwal98@gmail.com},\, \textbf{Apurba Samanta}$^{1}$ \footnote{apurbasamanta79@gmail.com} \\[0.5em]
    $^{1}$School of Physical Sciences, Indian Institute of Technology Mandi, \\
    Mandi, 175005, Himachal Pradesh, India \\[0.5em]
    }
    \normalsize

\date{} 

\begin{document}

\maketitle

\begin{abstract}
Cosmological $\alpha$-attractor models are renowned for their universal predictions for inflationary observables, largely independent of the detailed form of the inflaton potential. In this work, we investigate the extent to which this universality persists for polynomial $\alpha$-attractor models,
with particular emphasis on the scalar spectral index ($n_s$) and the post-inflationary reheating history. We derive an extended analytical expression for the scalar spectral index ($n_s$) and the tensor-to-scalar ratio ($r$), retaining sub-leading contributions beyond the commonly used universal approximation. Interestingly, we find that these corrections introduce an explicit dependence of $n_s$ on the parameter characterizing the polynomial attractor potential $\mu$, a dependence that is absent in the leading universal expression. We systematically investigate how this model dependence modifies the predictions in the $(n_s-r)$ plane and affects the  model parameter ($\mu$). We further incorporate the Duration of reheating ($\Delta N_{rh}$) and effective equation of state parameter of the reheating epoch ($w_{rh}$), accounting for their impact on the number of e-folds between horizon crossing and the end of inflation ($\Delta N_{\mathrm{CMB}}$). By confronting the resulting predictions with current CMB observations, we identify the viable parameter space and examine the interplay between the model parameter ($\mu$) and reheating. Our results demonstrate that sub-leading corrections can provide a window into the underlying structure of polynomial attractors, offering a possible way to distinguish different members of the attractor class observationally.
\end{abstract}


\noindent

\section{Introduction}
\label{sec:intro}
Inflation \cite{PhysRevD.23.347, LINDE1982389, Linde:1983gd, Linde_1994, Starobinsky:1980te,Lemoine:2008zz} represents a proposed stage in the early evolution of the Universe, characterized by a period of accelerated, nearly exponential expansion. When incorporated into the standard Big Bang framework, inflation provides solutions to several fundamental cosmological issues, including the horizon and flatness problems. Moreover, inflation provides a natural framework for generating primordial quantum fluctuations, which later manifest as temperature perturbations in the CMB and serve as the initial seeds for the formation of large-scale cosmic structures. A wide range of inflationary models have been developed, and many of them remain compatible with current observational constraints \cite{2020,2014,2016,Martin_2014,Martin:2024qnn,MARTIN2024101653}.
Among the theoretically and phenomenologically interesting classes of inflationary models are plateau and attractor models \cite{PhysRevLett.114.141302}. A characteristic feature of such models is that their predictions for the scalar-spectral-index and tensor-to-scalar ratio can become relatively insensitive to the detailed form of the inflaton potential. This apparent universality is particularly prominent in the $\alpha$-attractor framework \cite{Kallosh:2015lwa,kallosh2025singularalphaattractors,kallosh2026uas}, where the geometry of the inflaton field space plays a central role in determining the inflationary dynamics. As a result, apparently different potentials can approach similar predictions in the large-$N$ limit, providing an appealing connection between theoretical model building and observational constraints.

Polynomial attractor models \cite{Kallosh:2022feu} constitute an interesting realization of this broader attractor picture. In contrast to models whose plateau behavior is specified solely through an asymptotic exponential or hyperbolic structure, polynomial-attractor potentials possess a characteristic parameter that controls the transition between different regimes of the potential. 
Although polynomial $\alpha$-attractor models have been extensively investigated in inflationary phenomenology and CMB analyses, the interplay between finite model-parameter corrections and the post-inflationary reheating history remains worthy of further study. Marciniak et al. \cite{Marcini} investigated the P-model class for a range of polynomial powers, incorporating inflaton decay and fragmentation during reheating and constraining the reheating temperature using cosmological observations. Bhattacharya et al. \cite{Bhattacharya:2022akq} studied exponential and polynomial $\alpha$-attractor models and obtained constraints on their model parameters from CMB data, but did not specifically investigate the combined effect of extended model-parameter-dependent expressions and reheating on the parameter space of the polynomial-attractor potential considered here . Iacconi et al. \cite{Iacconi:2023mnw} derived improved analytical expressions for $\alpha$-attractor observables by retaining finite-parameter corrections and reheating effects, with their analysis focused primarily on the T-model. Although the leading-order predictions of attractor models often exhibit a universal form, subleading corrections can retain information about the underlying potential; in particular, the scalar spectral index can acquire explicit dependence on the model parameter even when its leading-order expression is universal \cite{Iacconi:2023mnw}. Moreover, the number of e-folds between the horizon exit of the pivot scale and the end of inflation depends on the post-inflationary expansion history and hence on reheating \cite{PhysRevLett.113.041302,PhysRevD.82.023511}. A consistent determination of the model parameter space therefore requires these effects to be incorporated simultaneously.

In this work, we present a comprehensive investigation of polynomial-attractor models within a generalized framework. Our primary objective is to establish robust observational constraints on the model parameter $\mu$. This is motivated by Planck observations \cite{2020}, which indicate a strong preference for concave inflationary potentials over convex ones. Because the polynomial-attractor framework naturally accommodates both potential types (concave and convex), constraining $\mu$ enables us to isolate the specific subclass of models that remains phenomenologically viable. While an exhaustive analysis of all possible realizations is beyond the present scope, we restrict our focus to two representative model subsets to examine their phenomenology in detail. A key methodological improvement in this study is our departure from standard literature treatments, which typically rely on approximate expressions for the scalar spectral index ($n_s$) and the tensor-to-scalar ratio ($r$). Instead, we derive exact, extended expressions for these observables before taking appropriate analytical limits. Furthermore, motivated by recent findings that emphasize the critical role of the reheating epoch in modifying inflationary predictions \cite{Iacconi:2023mnw}, we explicitly incorporate non-instantaneous reheating dynamics into our framework. By leveraging these refined analytical expressions alongside the latest Cosmic Microwave Background (CMB) data, we systematically constrain both the parameter $\mu$ and the duration of reheating. Finally, we highlight the impact of our framework by contrasting the observational predictions derived from our exact formulation against those obtained using standard approximations.

\section{Polynomial-Attractor Model}
In this work, we consider the first-kind of polynomial-attractor potential, given by \cite{Kallosh:2022feu}
\begin{equation}\label{eq:Polynomial-Attractor Model}
V(\phi)=V_0\frac{\phi^{q}}{\phi^{q}+\mu^{q}},
\end{equation}
Here, we define the exponent as \(q=2n\) and consider the representative cases \(q=2,\,4,\,6,\) and so on. The corresponding behavior of the potential can be seen in Fig.~\ref{fig:poly_potent_alpha}. The potential exhibits a quadratic like behavior in the small-field regime, \(|\phi|\ll\mu\), whereas it asymptotically approaches a constant plateau, \(V(\phi)\simeq V_0\), in the large-field regime, \(|\phi|\gg\mu\).
\begin{center}
\includegraphics[width=0.7\linewidth]{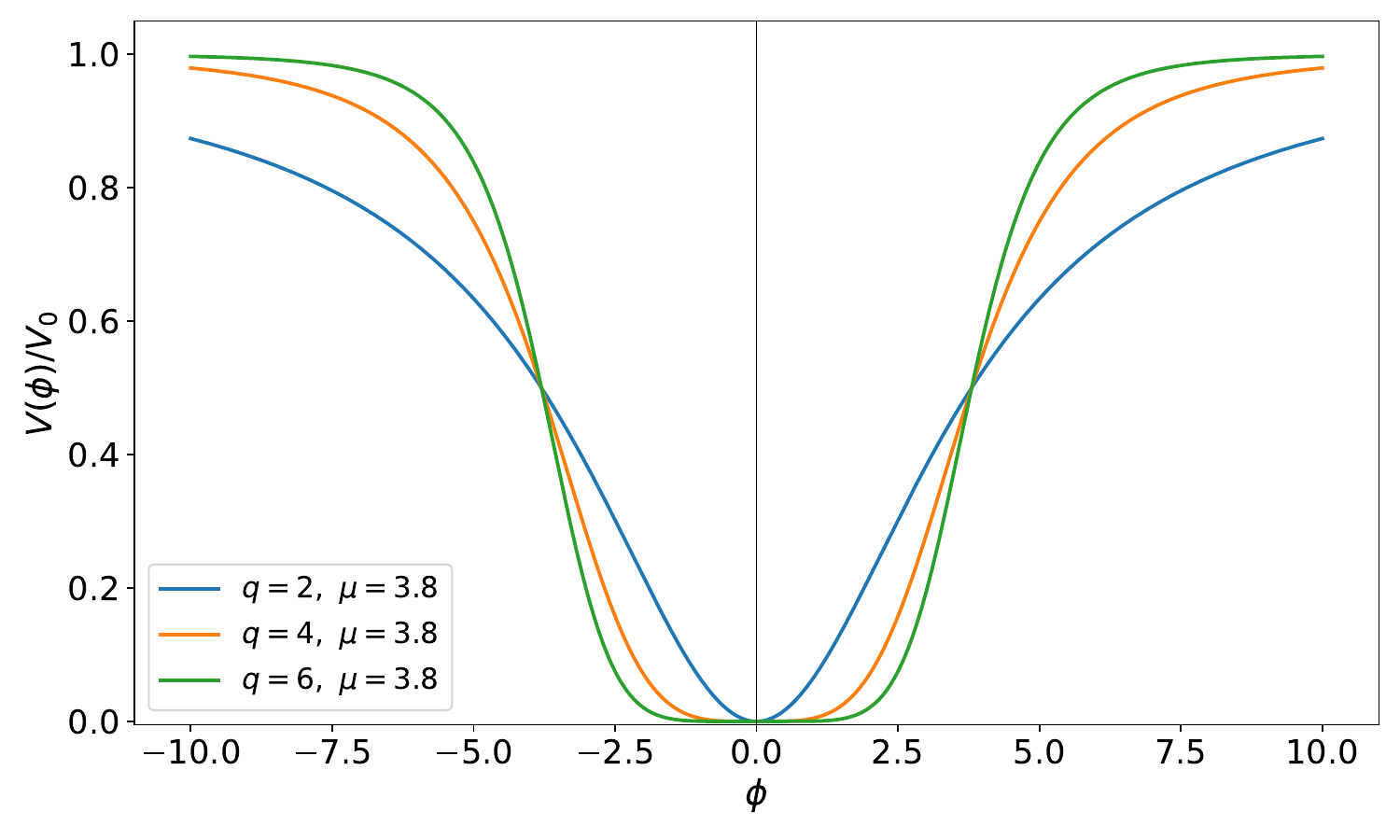}
\captionof{figure}{Polynomial-attractor models for different values of ($\mu$) and  fixed q=3.8, corresponding to distinct models, shown together in a single plot.}
\label{fig:poly_potent_alpha}
\end{center}

\subsection{General \(n_s\) and \(r\) for the P-Model without Reheating}
As established in the literature, the Planck 2018 observations \cite{2020} provide support for the applicability of the slow-roll approximation to inflationary dynamics. Within this framework, we obtain the general expressions for the potential slow-roll parameters associated with the potential given in Eq.~\eqref{eq:Polynomial-Attractor Model}. By integrating the corresponding slow-roll relation, the number of e-folds between the horizon-crossing field value \(\phi_{\mathrm{CMB}}\) and the end-of-inflation field value \(\phi_{\mathrm{end}}\) is given by
\begin{equation}\label{eq:delN _GEN_FORM}
\Delta N_\mathrm{CMB} \equiv N_\mathrm{end}-N_\mathrm{CMB}
\simeq \int_{\phi_\mathrm{end}}^{\phi_\mathrm{CMB}}
\mathrm{d}\phi\frac{V}{V_\phi}.
\end{equation}
Here, \(V_{\phi}\equiv dV/d\phi\) denotes the derivative of the potential with respect to the inflaton field \(\phi\). For the polynomial attractor potential given in Eq.~\eqref{eq:Polynomial-Attractor Model}, the above relation in Eq.~\eqref{eq:delN _GEN_FORM} reduces to
\begin{equation}\label{eq:gen_efold}
    \Delta N_{CMB}=\frac{1}{(q)(q+2)\mu^q}\left(\phi_{CMB}^{q+2}-\phi_{end}^{q+2}\right)
\end{equation}
At leading order, the end of inflation is conventionally determined by the condition \(\epsilon_V=1\). However, when second-order corrections are taken into account, the corresponding condition is modified and can be expressed as \cite{Ellis:2015pla}
\begin{equation}\label{eq:2nd order efsi}
\epsilon_V\simeq \left(1+\sqrt{1-\eta_V/2}\right)^2
\end{equation}
which provides the condition for the termination of inflation. The generalized expressions for the potential slow-roll parameters are given by
 \begin{equation}\label{eq:gen_1st para}
     \epsilon_V=\frac{1}{2}\left(\frac{V'}{V}\right)^2=\frac{1}{2}\left(\frac{q^2\mu^{2q}}{\phi^{2+2q}}\right)
 \end{equation}
 \begin{equation}\label{eq:gen_2nd para}
     \eta_V=\left(\frac{V''}{V}\right)=-q(q+1)\mu^q\phi^{-q-2}
 \end{equation}
We can also express Eqs.~\eqref{eq:gen_1st para} and \eqref{eq:gen_2nd para} in terms of \(\Delta N_{\mathrm{CMB}}\). Utilizing Eq.~\eqref{eq:gen_efold} reduces these expressions to Eqs.~\eqref{eq:poly_epsilon} and \eqref{eq:poly_eta}; a different form of this result appears in \cite{Kubota:2023ked}.
\begin{equation}\label{eq:poly_epsilon}
    \epsilon_V=\frac{q^2\mu^{2q}}{2\left(q(q+2)\mu^q\Delta N_{CMB}+\left(\frac{q \mu^q}{\sqrt 2}\right)^{\frac{q+2}{q+1}}\right)^{2(q+1)/q+2}}
\end{equation}
\begin{equation}\label{eq:poly_eta}
     \eta_V=\frac{-q(q+1)\mu^q}{q(q+2)\mu^q \Delta N_{CMB} +\left(\frac{\mu^{q} q}{\sqrt 2}\right)^{\frac{q+2}{q+1}}}
 \end{equation}
After using the equations \eqref{eq:gen_1st para} and \eqref{eq:gen_2nd para} into the equation \eqref{eq:2nd order efsi} we get the exact equation 
\begin{equation}\label{eq:exactequation}
    \phi^{2q+2}+\frac{3q(q+1)\mu^{q}}{4}\phi^{q}-\frac{q^{2}\mu^{2q}}{8}=0
\end{equation}
The above equation, Eq.~\eqref{eq:exactequation}, can be solved for specific values of \(q\), such as \(q=2,\,4,\,6,\ldots\). Solving the resulting equation for \(\phi\) gives the field value at which inflation terminates. We identify this solution as the end-of-inflation value, \(\phi_{\mathrm{end}}\). To establish a connection between the theoretical predictions and observational constraints, we consider the scalar power spectrum. Within the slow-roll approximation, the power spectrum can be expressed in terms of the inflaton potential as \cite{PhysRevD.110.030001} 
\begin{equation}
\label{power spectrum potentia}
P_\zeta(k)=\frac{V}{24\pi^2\epsilon_V}\Big|_{k=aH}.
\end{equation}
 For the observational analysis, we parameterize the primordial scalar power spectrum on large scales by a simple power-law form, given by \cite{2020}
\begin{equation}
\label{A_S_ power law}
    P_\zeta(k)=\mathcal{A}_s \left(\frac{k}{k_\text{CMB}} \right)^{n_s-1} \;,
\end{equation}
where $\mathcal{A}_s$ and $n_s$ are the amplitude and spectral tilt of $P_\zeta(k)$ respectively, defined at the scale $k_\text{CMB}=0.05\,\text{Mpc}^{-1}$. Equating Eqs.~\eqref{power spectrum potentia} and \eqref{A_S_ power law} at the pivot scale \(k=k_{\mathrm{CMB}}\),we obtain the normalization condition for the potential. Substituting the expression for \(\epsilon_V\) from Eq.~\eqref{eq:gen_1st para} evaluated at \(\phi=\phi_{\mathrm{CMB}}\), this relation takes the form
\begin{equation}\label{eq:power_relate_the}
\frac{V_0 \phi_{\mathrm{CMB}}^q}
{24\pi^2\left(\phi_{\mathrm{CMB}}^q+\mu^q\right)
\left(\dfrac{q^2\mu^{2q}}
{2\phi_{\mathrm{CMB}}^{2+2q}}\right)}
=
2.1\times10^{-9}.
\end{equation}
By substituting \(\phi_{\mathrm{CMB}}\) from Eq.~\eqref{eq:gen_efold}, and adopting the Planck value \(A_s=2.1\times10^{-9}\) \cite{2020}, expressed in terms of \(\Delta N_{\mathrm{CMB}}\) and \(\phi_{\mathrm{end}}\), into Eq.~\eqref{eq:power_relate_the}, and solving for \(V_0\), we obtain the corresponding expression for the potential parameter \(V_0\), which takes the form:
\begin{equation}\label{eq:V_{0}_EQUATION}
V_{0}=2.48714\times10^{-7}\,
\frac{
q^2\mu^{2q}
\left[
\mu^q+
\left(
\phi_{\mathrm{end}}^{q+2}
+\Delta N_{\mathrm{CMB}}\,q(q+2)\mu^q
\right)^{q/(q+2)}
\right]
}{
\left(
\phi_{\mathrm{end}}^{q+2}
+\Delta N_{\mathrm{CMB}}\,q(q+2)\mu^q
\right)^{(3q+2)/(q+2)}
}.
\end{equation}
For the polynomial attractor potential in Eq.~\eqref{eq:Polynomial-Attractor Model}, the scalar spectral index \(n_s\) and the tensor-to-scalar ratio \(r\) can be derived in their most general forms. Retaining the corrections from the model parameters and without assuming the large-\(\Delta N_{\mathrm{CMB}}\) limit, we obtain the following extended or non-universal expressions:
\begin{equation}\label{eq:poly_gen_ns}
\begin{aligned}
n_s = 1
- \frac{3q^2\mu^{2q}}
{\left[
q(q+2)\mu^q\Delta N_{CMB}
+\left(\frac{q\mu^q}{\sqrt{2}}\right)^{\frac{q+2}{q+1}}
\right]^{\frac{2(q+1)}{q+2}}} 
- \frac{2q(q+1)\mu^q}
{q(q+2)\mu^q\Delta N_{CMB}
+\left(\frac{q\mu^q}{\sqrt{2}}\right)^{\frac{q+2}{q+1}}}.
\end{aligned}
\end{equation}
\begin{equation}\label{eq:poly_gen_r}
   r=\frac{8q^2\mu^{2q}}{\left(q(q+2)\mu^q\Delta N_{CMB}+\left(\frac{q\mu^q}{\sqrt 2}\right)^{\frac{q+2}{q+1}}\right)^{2(q+1)/q+2}}\\  
\end{equation}
In the large-\(\Delta N_{\mathrm{CMB}}\) limit, the above equations, \eqref{eq:poly_gen_ns} and \eqref{eq:poly_gen_r}, can be simplified to recover the standard universal relations \cite{Kallosh:2022feu,Bhattacharya:2022akq}:
\begin{equation}\label{eq:simple_ns}
    n_s \approx1-\frac{2(q+1)}{(q+2) \Delta N_{CMB} }
\end{equation}
\begin{equation}\label{eq:simple_r}
    r \approx \frac{8q^2\mu^{\frac{2q}{q+2}}}{\left(q(q+2)\Delta N_{CMB}\right)^{\frac{2q+2}{q+2}}}
\end{equation}
Up to this point, we have obtained all potential and slow-roll parameters, as well as key quantities such as $V_0$ and $\phi_{\text{end}}$, in a general formulation for $q = 2n$ independent of reheating. Building upon this, the subsequent section extends the $q = 2n$ framework to evaluate all parameters associated with the reheating epoch.
\section{Reheating}
The reheating epoch marks the critical transition from inflation to the standard thermal universe, during which the inflaton field undergoes rapid coherent oscillations about its potential minimum, transferring its energy density into relativistic degrees of freedom. For an inflaton potential exhibiting an asymptotic power-law behavior, $V(\phi) \propto \phi^q$, in the vicinity of this minimum, the time-averaged effective equation-of-state parameter during the oscillatory phase is given by \cite{PhysRevD.28.1243}
\begin{equation}\label{eq:eq_state}
    w_{rh} = \frac{q-2}{q+2}.
\end{equation}
This scaling relation is directly applicable to the polynomial attractor models investigated in this work. Consequently, the parameter $w_{rh}$ governs the background expansion dynamics throughout the reheating phase, recovering a matter-dominated ($w_{rh} = 0$) or radiation-dominated ($w_{rh} = 1/3$) expansion history for quadratic and quartic minima, respectively. Furthermore, adopting the theoretical framework, we consider the limit of instantaneous reheating, parameterized by $\Delta N_{\mathrm{CMB,ir}}$, which is actually $\Delta N_{CMB}$ only when no extended reheating phase occurs, which can be confirmed from later equations for $\Delta N_{rh}=0$ (Number of reheating e-folds=0). This scenario posits an abrupt transition to the thermal era, thereby bypassing a prolonged post-inflationary reheating phase.We first start by assuming instant reheating itself ($\Delta N_{rh}=0$) but later reintroduce extended reheating phase $\Delta N_{rh} \neq 0$. By incorporating present-day cosmological parameters, the cosmic microwave background (CMB) pivot scale, and the effective relativistic degrees of freedom, the expression for $\Delta N_{\mathrm{CMB,ir}}$ takes the following form \cite{Martin:2013nzq, Iacconi:2023mnw}:
\begin{equation}
\label{Nstar step1}
    \Delta N_{\mathrm{CMB,ir}}\simeq 61.02 +\frac{1}{4}\ln{\left(\frac{V_\text{CMB}^2}{ \rho_\text{end}} \right) } \;,
\end{equation}
Here, \(\rho_{\mathrm{end}}\) is determined from the Friedmann constraint by imposing the condition \(\dot{\phi}_{\mathrm{end}}^{\,2}/2 = H_{\mathrm{end}}^{2}\) at the end of inflation. For the polynomial attractor model, it is given by the following expression:
\begin{equation}\label{eq:roh end equation}
    \rho_\text{end}=\frac{1}{2}{\dot{\phi}_\text{end}}^2 +V(\phi_\text{end}) = \frac{3}{2}V(\phi_\text{end}) =\frac{3}{2}V_0\Big[\frac{\phi_{end}^q}{\phi_{end}^q+\mu^q}\Big],
\end{equation}
Equation \eqref{Nstar step1} can be expressed in terms of \(V_{\mathrm{end}}\) by substituting \(\rho_{\mathrm{end}}\) using Eq.~\eqref{eq:roh end equation}. This substitution leads to the following form:
\begin{equation}
\label{Nstar step2}
    \Delta N_\text{CMB,ir}\simeq 60.92 +\frac{1}{4}\ln{\left(\frac{V_\text{CMB}^2}{ V_\text{end}} \right) } \;.
\end{equation}
To express Eq.~\eqref{Nstar step2} in terms of \(\phi_{\mathrm{end}}\), we substitute the general potential given in Eq.~\eqref{eq:Polynomial-Attractor Model} and \eqref{eq:V_{0}_EQUATION} into Eq.~\eqref{Nstar step2}. This allows us to rewrite the equation in terms of the end-of-inflation field value \(\phi_{\mathrm{end}}\), yielding the following modified expression:
\begin{align}\label{eq:Nstar step3}
\Delta N_{\mathrm{CMB,ir}}
&=
57.11826
+
\frac{1}{4}
\ln\left[
\frac{\mu^{2q}q^2g}{f}
\left(
\frac{\phi_{\mathrm{end}}^q+\mu^q}
{\phi_{\mathrm{end}}^q}
\right)
\right]
\\[6pt]
g
&=
\left[
\phi_{\mathrm{end}}^{q+2}
+
\Delta N_{\mathrm{CMB,ir}}(q+2)q\mu^q
\right]^{-1} \notag
\\[6pt]
f
&=
\mu^q
+
g^{-\frac{q}{q+2}}.\notag
\end{align}
The equation \eqref{eq:Nstar step3} can be solved exactly with the aid of Eq.~\eqref{eq:exactequation}. By substituting the relevant values of \(q\) into Eq.~\eqref{eq:exactequation}, we obtain the corresponding value of \(\phi_{\mathrm{end}}\). In the subsequent section, we consider two specific cases and solve the resulting equations separately for each case.

In the preceding discussion, we introduce the reheating phase following inflation. The duration of reheating, characterized by the number of reheating e-folds \(\Delta N_{\mathrm{rh}}\), is defined as the interval between the end of inflation and the completion of reheating. The corresponding energy densities at these two stages are denoted by \(\rho_{\mathrm{end}}\) and \(\rho_{\mathrm{th}}\), respectively.
 \begin{equation}
    \label{eq:N_rh equation}
    \Delta N_\text{rh} \equiv \frac{1}{3(1+w_{rh})}\log{\left( \frac{\rho_\text{end}}{\rho_\text{th}}\right)} \;.
\end{equation}
The relation connecting the reheating duration \(\Delta N_{\mathrm{rh}}\) with the energy densities at the end of inflation and the completion of reheating can be expressed in terms of the effective equation-of-state parameter \(w_{\mathrm{rh}}\), as defined in Eq.~\eqref{eq:eq_state}. Thus, \(\Delta N_{\mathrm{rh}}\) depends on both \(w_{\mathrm{rh}}\) and the energy density \(\rho_{\mathrm{th}}\). At the onset of Big-Bang nucleosynthesis (BBN), we take the energy density to be \(\rho_{\mathrm{th}}=(1\,\mathrm{TeV})^4\), where \(1\,\mathrm{TeV}=10^3\,\mathrm{GeV}\) and the reduced Planck mass is \(M_{\mathrm{pl}}=2.435\times10^{18}\,\mathrm{GeV}\). We adopt these values throughout the analysis presented in this paper. By requiring that reheating is completed no later than the onset of BBN, we obtain the maximum allowed duration of the reheating phase. Consequently, the maximum allowed number of reheating e-folds is given by:
\begin{equation}
\label{eq:N_rh_max_eq}
\Delta N_{\mathrm{rh}}
\leq
\frac{1}{3(1+w_{\mathrm{rh}})}
\log\left(
\frac{\rho_{\mathrm{end}}}{(1\,\mathrm{TeV})^4}
\right)
\equiv
\Delta N_{\mathrm{rh,max}}.
\end{equation}
This relation determines the upper bound on the duration of the reheating phase for a given equation-of-state parameter \(w_{\mathrm{rh}}\). Accordingly, throughout the analysis, we consider the reheating e-folds to lie within the range $\Delta N_{\mathrm{rh}}\in[0,\Delta N_{\mathrm{rh,max}}]$,
This range is adopted for a general \(w_{\mathrm{rh}}\) and will also be applied to the specific model cases discussed in the subsequent sections.

Combining the results obtained above, the total number of CMB e-folds, \(\Delta N_{CMB}(\mu,\Delta N_{rh})\), can be expressed in terms of the instantaneous-reheating contribution, \(\Delta N_{\mathrm{CMB,ir}}\), and the additional contribution from the reheating phase, \(\Delta N_{\mathrm{rh}}\). The resulting relation can be written as follows:
\begin{equation}
    \Delta N_{CMB}(\mu,\Delta N_{rh})= \Delta N_{CMB, ir}(\mu)-\frac{1-3w_{rh}}{4} \Delta N_\text{rh}
    \label{eq:inclue_inst_rh}
\end{equation}
However, the general e-fold relation given in Eq.~\eqref{eq:gen_efold} does not contain any explicit dependence on \(\Delta N_{\mathrm{rh}}\). In the absence of reheating effects, the total number of CMB e-folds, \(\Delta N_{\mathrm{CMB}}\), is determined solely by the inflaton field values and the model parameter \(\mu\). On the other hand, once the effects of the reheating phase are taken into account, Eq.~\eqref{eq:inclue_inst_rh} shows that \(\Delta N_{\mathrm{CMB}}\) acquires an explicit dependence on both \(\mu\) and the number of reheating e-folds, \(\Delta N_{\mathrm{rh}}\). Therefore, in order to incorporate the reheating effects consistently into the inflationary observables, the CMB e-fold number appearing in the expressions for the scalar spectral index and tensor-to-scalar ratio must be replaced by the reheating-dependent quantity \(\Delta N_{\mathrm{CMB}}(\mu,\Delta N_{\mathrm{rh}})\). Consequently, the expressions for the scalar spectral index and tensor-to-scalar ratio, given in Eqs.~\eqref{eq:poly_gen_ns} and \eqref{eq:poly_gen_r}, respectively, can be rewritten in terms of \(\Delta N_{\mathrm{CMB}}(\mu,\Delta N_{\mathrm{rh}})\). The resulting expressions take the improved form:
\begin{equation}\label{eq:re_poly_gen_ns}
\begin{aligned}
n_s = 1
- \frac{3q^2\mu^{2q}}
{\left[
q(q+2)\mu^q\Delta N_{CMB}(\mu,\Delta N_{rh})
+\left(\frac{q\mu^q}{\sqrt{2}}\right)^{\frac{q+2}{q+1}}
\right]^{\frac{2(q+1)}{q+2}}}\\ 
- \frac{2q(q+1)\mu^q}
{q(q+2)\mu^q\Delta N_{CMB}(\mu,\Delta N_{rh})
+\left(\frac{q\mu^q}{\sqrt{2}}\right)^{\frac{q+2}{q+1}}}.
\end{aligned}
\end{equation}
\begin{equation}\label{eq:re_poly_gen_r}
   r=\frac{8q^2\mu^{2q}}{\left[q(q+2)\mu^q\Delta N_{CMB}(\mu,\Delta N_{rh})+\left(\frac{q\mu^q}{\sqrt 2}\right)^{\frac{q+2}{q+1}}\right]^{2(q+1)/q+2}}  
\end{equation}
In the large-\(\Delta N_{\mathrm{CMB}}\) limit, the above equations, \eqref{eq:re_poly_gen_ns} and \eqref{eq:re_poly_gen_r}, can be simplified to recover the standard relations:
\begin{equation}\label{eq:sim_re_ns}
    n_s \approx1-\frac{2(q+1)}{(q+2) \Delta N_{CMB}(\mu,\Delta N_{rh}) }
\end{equation}
\begin{equation}\label{eq:sim_re_r}
    r \approx \frac{8q^2\mu^{\frac{2q}{q+2}}}{[(q)(q+2)\Delta N_{CMB}(\mu,\Delta N_{rh})]^{\frac{2q+2}{q+2}}}
\end{equation}
At this stage, we have developed the general framework for the case \(q=2n\). To investigate the model in greater detail, it is necessary to determine the relevant parameters explicitly and examine their behaviour across the parameter space. While the formalism is applicable to arbitrary even values of \(q\), we focus our detailed analysis on two representative cases, \(q=2\) and \(q=4\). The explicit expressions for the relevant parameters and their corresponding behaviour are derived and discussed in detail in the subsequent sections.
\section{Methods and detailed Investigation of the Models}
\subsection{Case 1: q=2}
\begin{figure}[htbp]
    \centering

    \begin{subfigure}[b]{0.48\textwidth}
        \centering
        \includegraphics[width=\linewidth]{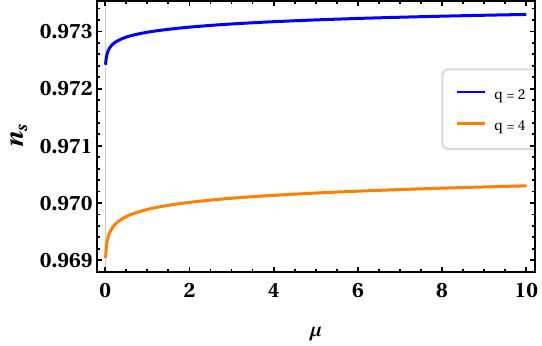}
    \end{subfigure}
    \hfill
    \begin{subfigure}[b]{0.48\textwidth}
        \centering
        \includegraphics[width=\linewidth]{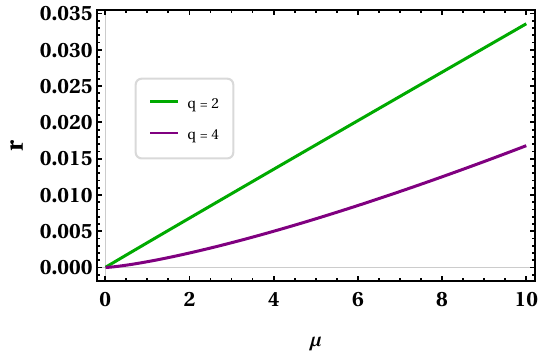}
    \end{subfigure}
    \caption{Comparison of the universal behavior of the scalar spectral index $n_s$ and the tensor-to-scalar ratio $r$ as functions of $\mu$, obtained from Eqs.~\eqref{eq:sim_1} for $q=2$ and \eqref{eq:sim_2} for $q=4$ for different models. }
    \label{fig:univer_ns_r}
\end{figure}

\begin{figure}[htbp]
    \centering

    \begin{subfigure}[b]{0.48\textwidth}
        \centering
        \includegraphics[width=\linewidth]{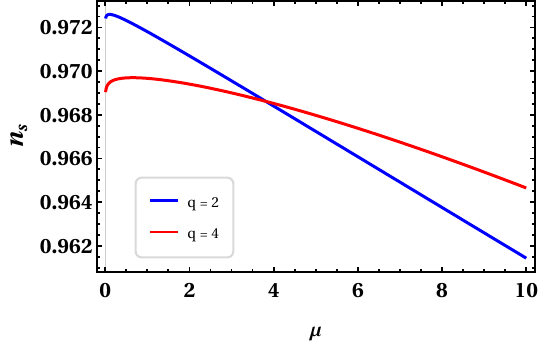}
    \end{subfigure}
    \hfill
    \begin{subfigure}[b]{0.48\textwidth}
        \centering
        \includegraphics[width=\linewidth]{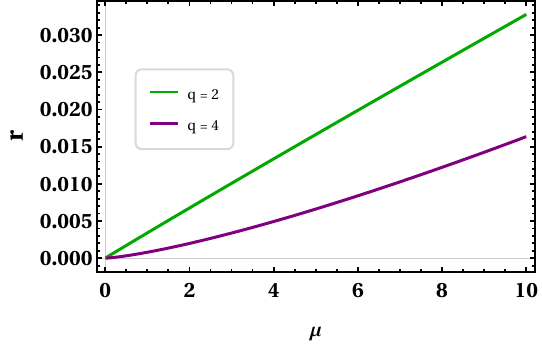}
    \end{subfigure}

    \caption{Comparison of the non-universal behavior of $n_s$ and $r$ as functions of $\mu$ for different models, obtained from Eqs.~\eqref{eq:sim_1st_ns}--\eqref{eq:sim_1st_r} for $q=2$ and \eqref{eq:sim_2nd_ns}--\eqref{eq:sim_2nd_r} for $q=4$.}
    \label{fig:non_univer_ns_r}
\end{figure}

\begin{figure}[htbp]
    \centering

    \begin{subfigure}[b]{0.49\textwidth}
        \centering
        \includegraphics[width=\linewidth]{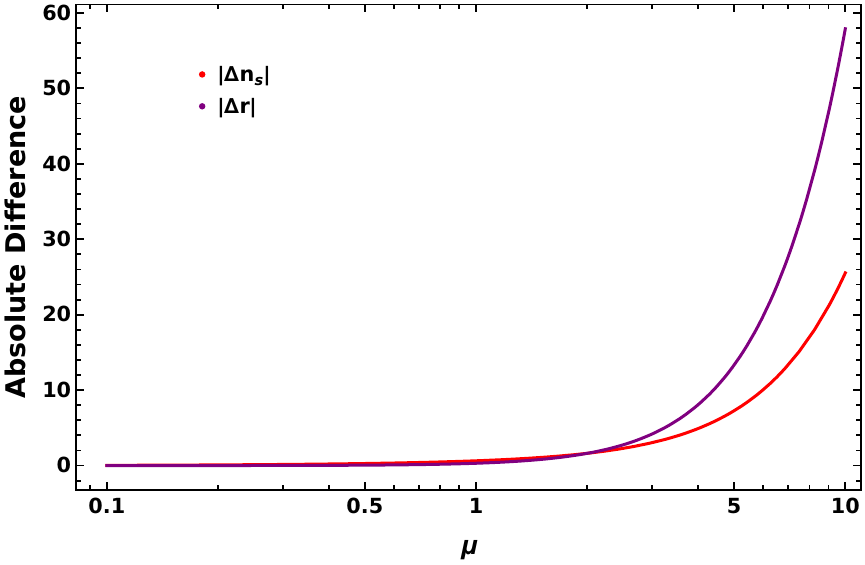}
        \caption{For q=2}
        \label{fig:q_2_absolute_diff_ns_r}
    \end{subfigure}
    \hfill
    \begin{subfigure}[b]{0.49\textwidth}
        \centering
        \includegraphics[width=\linewidth]{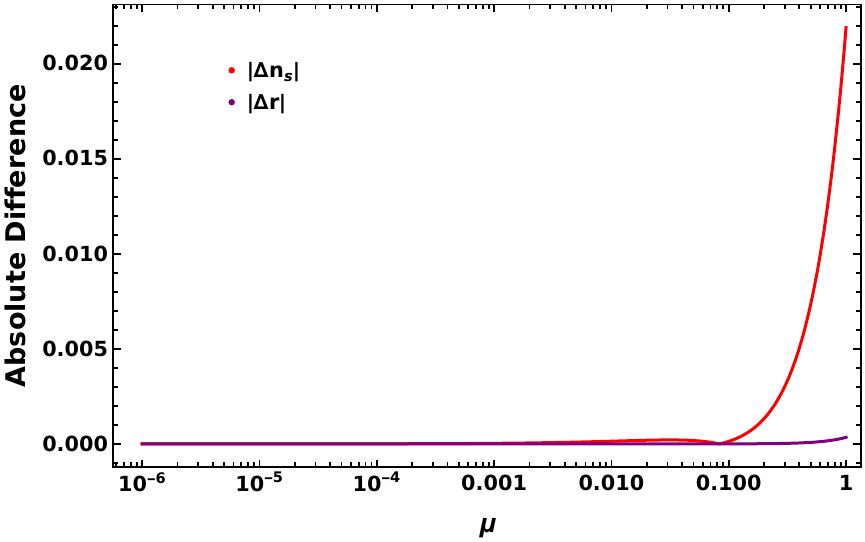}
        \caption{For q=4}
        \label{fig:q_4_absolute_diff_ns_r}
    \end{subfigure}
    \caption{Percentage deviation of the non-universal from the universal forms of $n_s$ and $r$ for different models, illustrating the deviations between the two cases. }
\end{figure}

\begin{figure}[htbp]
    \centering

    \begin{subfigure}[b]{0.48\textwidth}
        \centering
        \includegraphics[width=\linewidth]{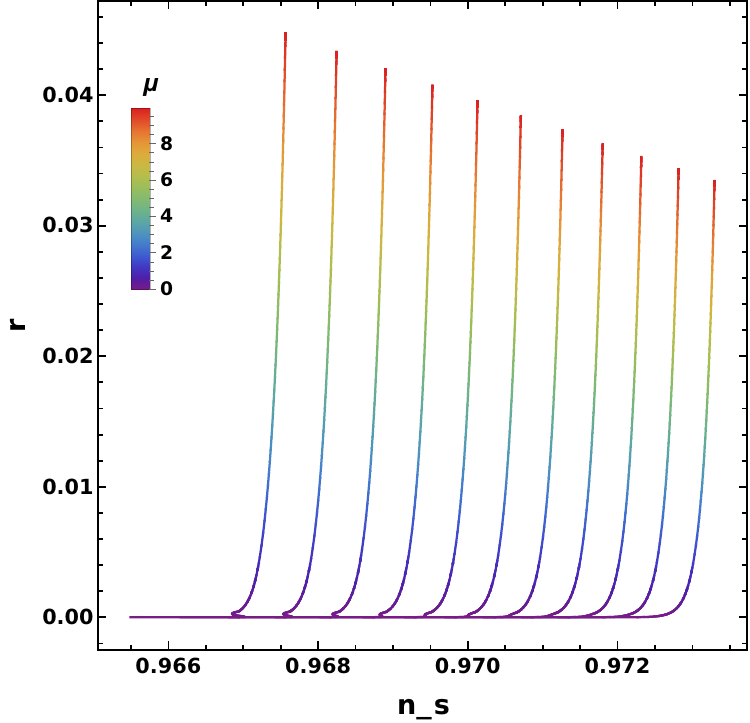}
    \end{subfigure}
    \hfill
    \begin{subfigure}[b]{0.48\textwidth}
        \centering
        \includegraphics[width=\linewidth]{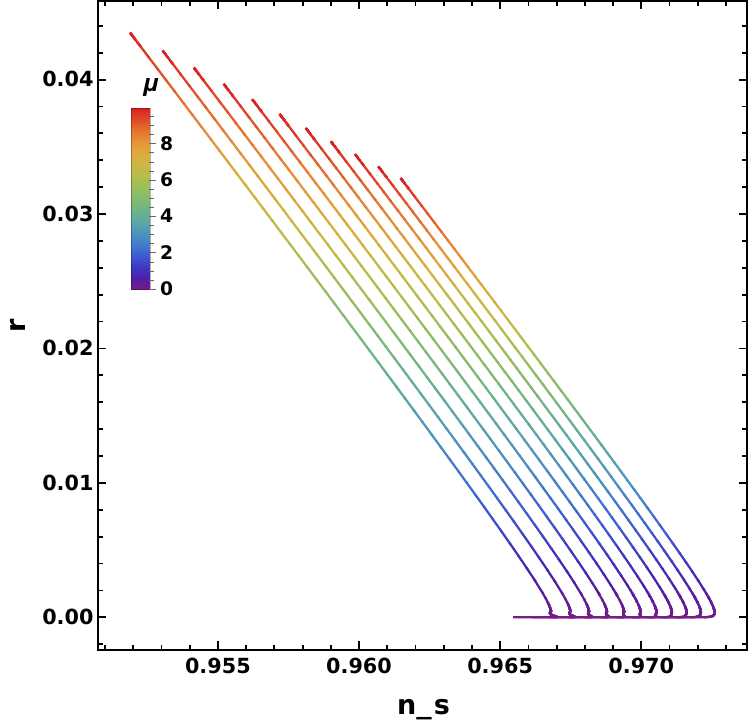}
    \end{subfigure}

    \caption{Variation of \(n_s\) and \(r\) with \(\mu\) for different values of \(\Delta N_{\mathrm{rh}}\) for the case \(q=2\), with \(\Delta N_{\mathrm{rh}}\) increasing from right to left. The left panel shows the universal approximations (Eqs.~\eqref{eq:sim_re_ns}--\eqref{eq:sim_re_r}), and the right panel displays the non-universal results (Eqs.~\eqref{eq:re_poly_gen_ns}--\eqref{eq:re_poly_gen_r}).   }
    \label{fig:univs nonuni}
\end{figure}

For the case \(q=2\), the scalar spectral index \(n_s\) and the tensor-to-scalar ratio \(r\) can be obtained directly from the general expressions given in Eqs.~\eqref{eq:poly_gen_ns} and \eqref{eq:poly_gen_r}. Substituting \(q=2\) into these generalized expressions, and considering Eq.~\eqref{eq:inclue_inst_rh}, we assume reheating, for which \(\Delta N_{\mathrm{rh}}=0\). Consequently, \(\Delta N_{\mathrm{CMB}}\) can be expressed in terms of \(\Delta N_{\mathrm{CMB,ir}}\), yielding
\begin{equation}\label{eq:sim_1st_ns}
    n_s=1- \frac{12\mu^{4}}{\left(8\mu^2\Delta N_{\mathrm{CMB,ir}}+({2\mu^4})^{\frac{2}{3}}\right)^{3/2}}-\frac{12\mu^2}{8\mu ^2 \Delta N_{\mathrm{CMB,ir}} +({2\mu^4})^{\frac{2}{3}}}
\end{equation}
\begin{equation}\label{eq:sim_1st_r}
    r=\frac{32\mu^4}{\left(8\mu^2\Delta N_{\mathrm{CMB,ir}}+(2\mu^4)^{2/3}\right)^{3/2}}
\end{equation}
Again, in the large-\(\Delta N_{\mathrm{CMB,ir}}\) limit, the two generalized results, Eqs.~\eqref{eq:simple_ns} and \eqref{eq:simple_r}, can be simplified to
\begin{equation}\label{eq:sim_1}
    n_s\approx 1-\frac{3}{2 \Delta N_{\mathrm{CMB,ir}} } ;\hspace{0.5cm} r\approx \frac{\sqrt2 \mu}{(\Delta N_{\mathrm{CMB,ir}})^{3/2}}
\end{equation}
Next, we calculate \(\phi_{\mathrm{end}}\) from Eq.~\eqref{eq:exactequation}. By substituting \(q=2\) into this equation, we obtain an expression for \(\phi_{\mathrm{end}}\) in terms of \(\mu\), which takes the following form:
\begin{equation}\label{eq:end for q=2}
    \phi_{end}= \frac{\sqrt{\mu}}{2^{1/3}}\left((\mu+ \sqrt{\mu^2+2})^{1/3}+(\mu- \sqrt{\mu^2+2})^{1/3}\right)^{1/2}
\end{equation}
Using the $\Delta N_{\mathrm{CMB,ir}}$ values obtained from Eq.~\eqref{eq:Nstar step3} for $q = 2$ along with their variation with $\mu$ shown in Fig. \ref{fig:dNcmb_ins_reh_plot}, we calculate the corresponding scalar spectral index ($n_s$) and tensor-to-scalar ratio ($r$) for the subsequent analysis.
\begin{center}
\includegraphics[width=0.5\linewidth]{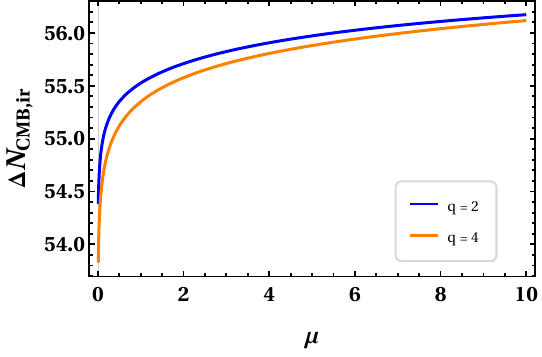}
\captionof{figure}{Variation of $\Delta N_{\mathrm{CMB,ir}}$ as a function of $\mu$, derived from equation \eqref{eq:Nstar step3} for $q = 2$ and $q = 4$.}
\label{fig:dNcmb_ins_reh_plot}
\end{center}
To incorporate the effects of reheating, we calculate the scalar spectral index ($n_s$) and the tensor-to-scalar ratio ($r$) for $q = 2$. The non-universal predictions, evaluated using Eqs.~\eqref{eq:sim_1st_ns} and \eqref{eq:sim_1st_r}, are illustrated in Fig.~\ref{fig:non_univer_ns_r}, whereas the corresponding universal forms given by Eqs.~\eqref{eq:sim_1} are presented in Fig.~\ref{fig:univer_ns_r}. Furthermore, Fig.~\ref{fig:q_2_absolute_diff_ns_r} highlights the absolute differences between the universal and non-universal formulations for both parameters.
From Eq.~\eqref{eq:N_rh_max_eq}, the value of $w_{rh}$ can be determined by setting $q=2$ in Eq.~\eqref{eq:eq_state}, which yields $w_{rh}=0$. Consequently, Eq.~\eqref{eq:N_rh_max_eq} can be rewritten in the following form:
\begin{equation}
\label{eq:new_N_rh_max_eq}
\Delta N_{\mathrm{rh}}
\leq
\frac{1}{3}
\log\left(
\frac{\rho_{\mathrm{end}}}{(1\,\mathrm{TeV})^4}
\right)
\equiv
\Delta N_{\mathrm{rh,max}}.
\end{equation}
\begin{figure}[htbp]
    \centering

    \begin{subfigure}[b]{0.49\textwidth}
        \centering
        \includegraphics[width=\linewidth]{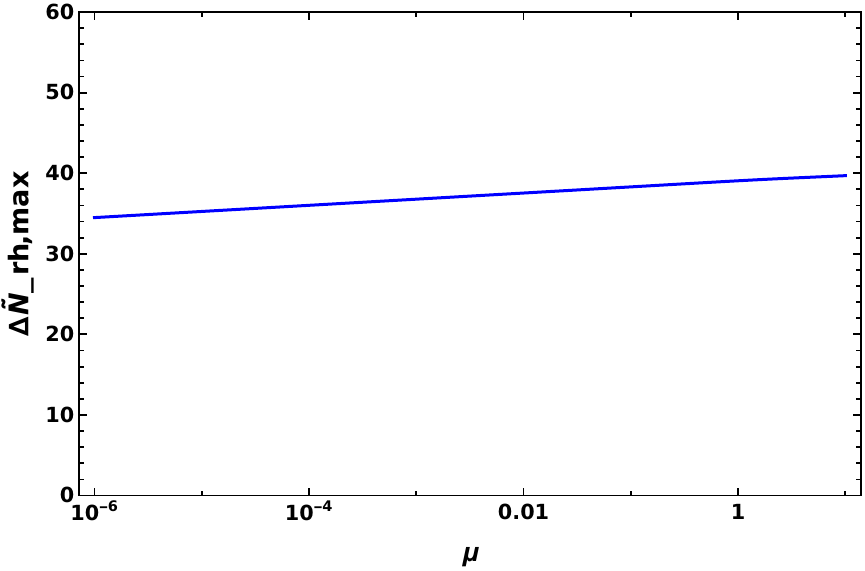}
        \caption{For q=2}
        \label{fig:q_2deltaNrhMax_plot}
    \end{subfigure}
    \hfill
    \begin{subfigure}[b]{0.49\textwidth}
        \centering
        \includegraphics[width=\linewidth]{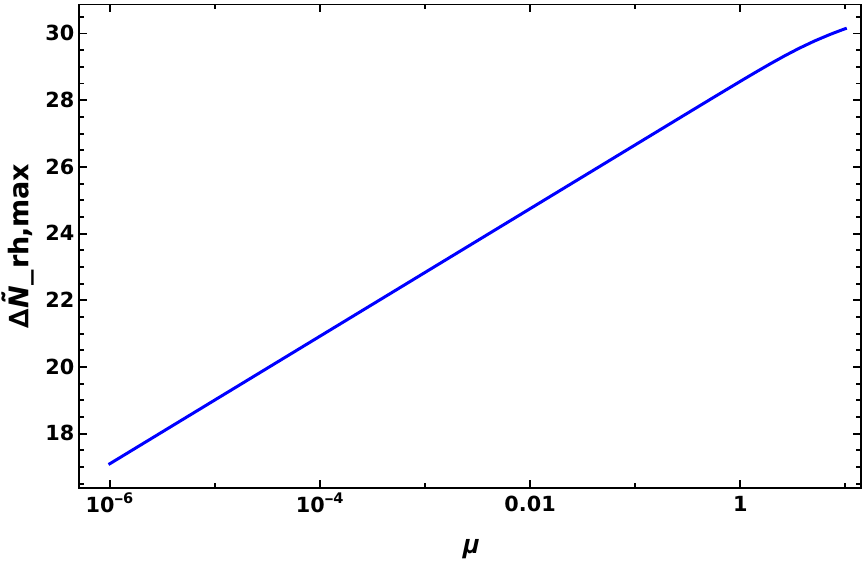}
        \caption{For q=4}
        \label{fig:q_4deltaNrhMax_plot}
    \end{subfigure}
    \caption{ Behavior of $\Delta N_{\mathrm{rh,max}}$ as a function of $\mu$, obtained from Eq.~\eqref{eq:new_N_rh_max_eq} for $q=2$ (a) and Eq.~\eqref{eq:q_4_new_N_rh_max_eq} for $q=4$ (b).}
\end{figure}
We also observe that \(\Delta N_{\mathrm{rh,max}}\) exhibits a dependence on the parameter \(\mu\), as illustrated in Fig.~\ref{fig:q_2deltaNrhMax_plot}. Having accounted for the reheating phase, we set $w_{\text{rh}} = 0$ for the $q = 2$ case in Eq.~\eqref{eq:inclue_inst_rh}, which simplifies the expression to
\begin{equation}
\Delta N_{\mathrm{CMB}}(\mu,\Delta N_{\mathrm{rh}}) = \Delta N_{\mathrm{CMB, ir}}(\mu) - \frac{1}{4} \Delta N_{\mathrm{rh}}.
\label{eq:rewrite_inclue_inst_rh}
\end{equation}
This relation clearly shows how the reheating duration ($\Delta N_{\mathrm{rh}}$) enters into $\Delta N_{\mathrm{CMB}}(\mu, \Delta N_{\mathrm{rh}})$. Using this, we calculate and plot the scalar spectral index ($n_s$) and tensor-to-scalar ratio ($r$) against $\mu$ and $\Delta N_{\mathrm{rh}}$. Both the generalized non-universal equations (Eqs.~\eqref{eq:re_poly_gen_ns}--\eqref{eq:re_poly_gen_r}) and the universal approximations (Eqs.~\eqref{eq:sim_re_ns}--\eqref{eq:sim_re_r}) are compared in Fig.~\ref{fig:univs nonuni}.
\subsection{Case 2: q=4}
For the case \(q=4\), the scalar spectral index \(n_s\) and the tensor-to-scalar ratio \(r\) can be obtained directly from the general expressions given in Eqs.~\eqref{eq:poly_gen_ns} and \eqref{eq:poly_gen_r}. Substituting \(q=4\) into these generalized expressions, and considering Eq.~\eqref{eq:inclue_inst_rh}, we assume reheating, for which \(\Delta N_{\mathrm{rh}}=0\). Consequently, \(\Delta N_{\mathrm{CMB}}\) can be expressed in terms of \(\Delta N_{\mathrm{CMB,ir}}\), yielding
\begin{equation}\label{eq:sim_2nd_ns}
    n_s=1- \frac{48\mu^{8}}{\left[24\mu^4\Delta N_{CMB,ir}+({2 \sqrt{2}\mu^4})^{\frac{6}{5}}\right]^{5/3}}-\frac{40\mu^4}{24\mu ^4 \Delta N_{CMB,ir} +({2\sqrt{2}\mu^4})^{\frac{6}{5}}}
\end{equation}
\begin{equation}\label{eq:sim_2nd_r}
    r=\frac{128\mu^8}{\left[24\mu^4\Delta N_{CMB,ir}+(2 \sqrt{2}\mu^4)^{6/5}\right]^{5/3}}
\end{equation}
Again, in the large-\(\Delta N_{\mathrm{CMB,ir}}\) limit, the two generalized results, Eqs.~\eqref{eq:simple_ns} and \eqref{eq:simple_r}, can be simplified to
\begin{equation}\label{eq:sim_2}
    n_s\approx1-\frac{5}{3 \Delta  N_{CMB,ir} } ;\hspace{0.5cm} r\sim\frac{4 \mu^{4/3}}{(3\Delta  N_{CMB,ir})^{5/3}}
\end{equation}
For \(q=4\), a simplified analytical expression for \(\phi_{\rm end}\), analogous to that obtained for \(q=2\) in Eq.~\eqref{eq:end for q=2}, cannot be derived. This is because Eq.~\eqref{eq:exactequation} does not admit a suitable analytical root for \(q>2\). We therefore determine \(\phi_{\rm end}\) numerically by solving Eq.~\eqref{eq:exactequation} over the parameter range \(10^{-6}\leq\mu\leq10\). The resulting numerical solution for \(\phi_{\rm end}\) is subsequently used to evaluate \(\Delta N_{\rm CMB,ir}\). The quantity \(\Delta N_{\rm CMB,ir}\) can be obtained from Eq.~\eqref{eq:Nstar step3} by setting \(q=4\). The resulting dependence and its behavior are illustrated in Fig.~\ref{fig:dNcmb_ins_reh_plot}. We calculate the corresponding scalar spectral index ($n_s$) and tensor-to-scalar ratio ($r$) for the subsequent analysis. To incorporate the effects of reheating, we calculate the scalar spectral index ($n_s$) and the tensor-to-scalar ratio ($r$) for $q = 4$. The non-universal predictions, evaluated using \eqref{eq:sim_2nd_ns}--\eqref{eq:sim_2nd_r}, are illustrated in Fig.~\ref{fig:non_univer_ns_r}, whereas the corresponding universal forms given by Eqs.~\eqref{eq:sim_2}are presented in Fig.~\ref{fig:univer_ns_r}. Furthermore, Fig.~\ref{fig:q_4_absolute_diff_ns_r} highlights the absolute differences between the universal and non-universal formulations for both parameters. From Eq.~\eqref{eq:N_rh_max_eq}, the value of $w_{rh}$ can be determined by setting $q=4$ in Eq.~\eqref{eq:eq_state}, which yields $w_{rh}=\frac{1}{3}$. Consequently, Eq.~\eqref{eq:N_rh_max_eq} can be rewritten in the following form:
\begin{equation}
\label{eq:q_4_new_N_rh_max_eq}
\Delta N_{\mathrm{rh}}
\leq
\frac{1}{4}
\log\left(
\frac{\rho_{\mathrm{end}}}{(1\,\mathrm{TeV})^4}
\right)
\equiv
\Delta N_{\mathrm{rh,max}}.
\end{equation}
We also observe that \(\Delta N_{\mathrm{rh,max}}\) exhibits a dependence on the parameter \(\mu\), as illustrated in Fig.~\ref{fig:q_4deltaNrhMax_plot}.

Having accounted for the reheating phase, we set $w_{\text{rh}} = \frac{1}{3}$ for the $q = 4$ case in Eq.~\eqref{eq:inclue_inst_rh}, which simplifies the expression to:
\begin{equation}
\Delta N_{\mathrm{CMB}}(\mu) = \Delta N_{\mathrm{CMB, ir}}(\mu) 
\label{eq:rewrite_inclue_not_inst_rh}
\end{equation}
\begin{figure}[htbp]
    \centering

    \begin{subfigure}[b]{0.48\textwidth}
        \centering
        \includegraphics[width=\linewidth]{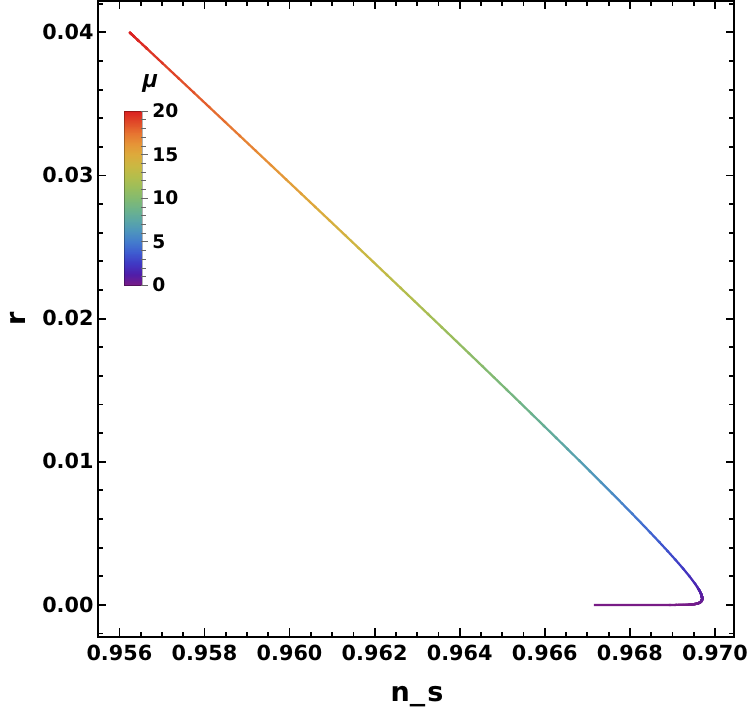}
    \end{subfigure}
    \hfill
    \begin{subfigure}[b]{0.48\textwidth}
        \centering
        \includegraphics[width=\linewidth]{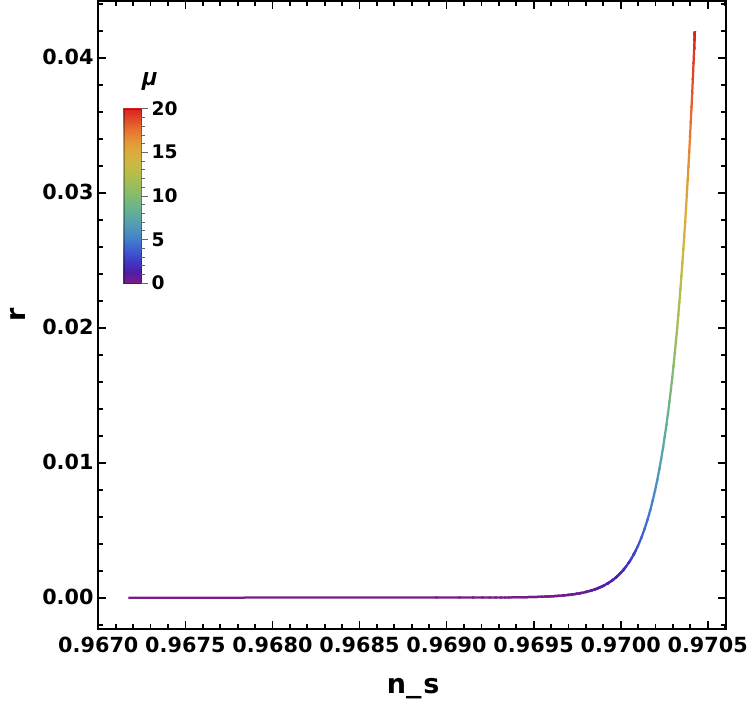}
    \end{subfigure}

    \caption{Variation of \(n_s\) and \(r\) with \(\mu\) for the case \(q=4\). As evident from the plot, \(n_s\) and \(r\) shows no dependence on \(\Delta N_{\mathrm{rh}}\). The right panel shows the universal approximations (Eqs.~\eqref{eq:sim_re_ns}--\eqref{eq:sim_re_r}), and the left panel displays the non-universal results (Eqs.~\eqref{eq:re_poly_gen_ns}--\eqref{eq:re_poly_gen_r}).   }
    \label{fig:q_4_univs nonuni}
\end{figure}
Equation~\eqref{eq:rewrite_inclue_not_inst_rh} explicitly demonstrates that \(\Delta N_{\rm CMB}\) is independent of the reheating contribution. Consequently, the analysis allows us to constrain \(\Delta N_{\rm CMB}\), but not \(\Delta N_{\rm rh}\). Using this relation, we evaluate and plot the scalar spectral index \(n_s\) and the tensor-to-scalar ratio \(r\) as functions of \(\mu\). We compare the results obtained from the generalized non-universal expressions, Eqs.~\eqref{eq:re_poly_gen_ns}--\eqref{eq:re_poly_gen_r}, with those from the corresponding universal approximations, Eqs.~\eqref{eq:sim_re_ns}--\eqref{eq:sim_re_r}, in Fig.~\ref{fig:q_4_univs nonuni}. Since \(\Delta N_{\rm rh}\) does not enter independently in this case, the resulting parameter space reduces to a single curve, parametrized solely by \(\mu\) and $\Delta N_{CMB}$.
\section{Discussion and Conclusions }
\begin{figure}[htbp]
    \centering

    \begin{subfigure}[b]{0.48\textwidth}
        \centering
        \includegraphics[width=\linewidth]{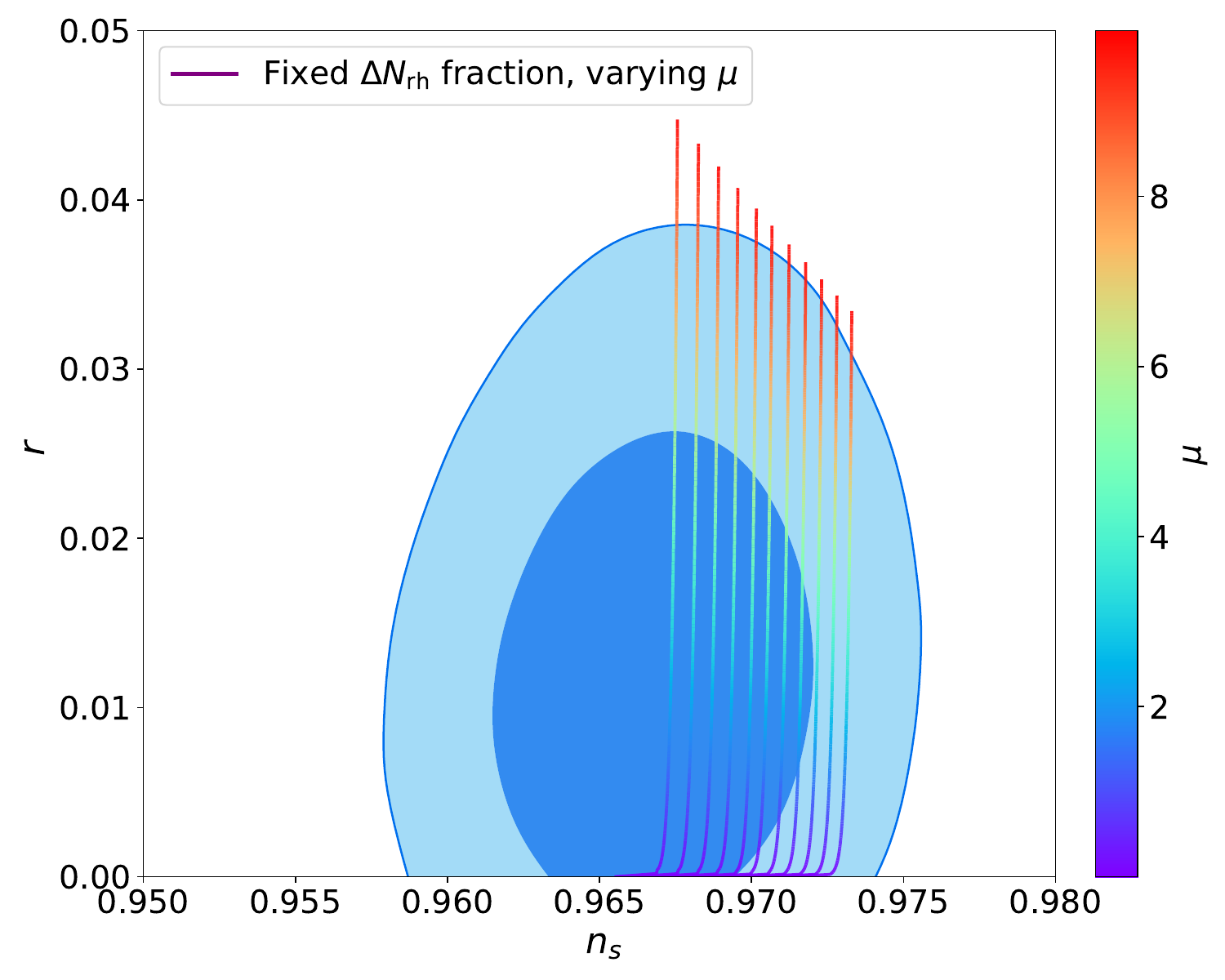}
    \end{subfigure}
    \hfill
    \begin{subfigure}[b]{0.48\textwidth}
        \centering
        \includegraphics[width=\linewidth]{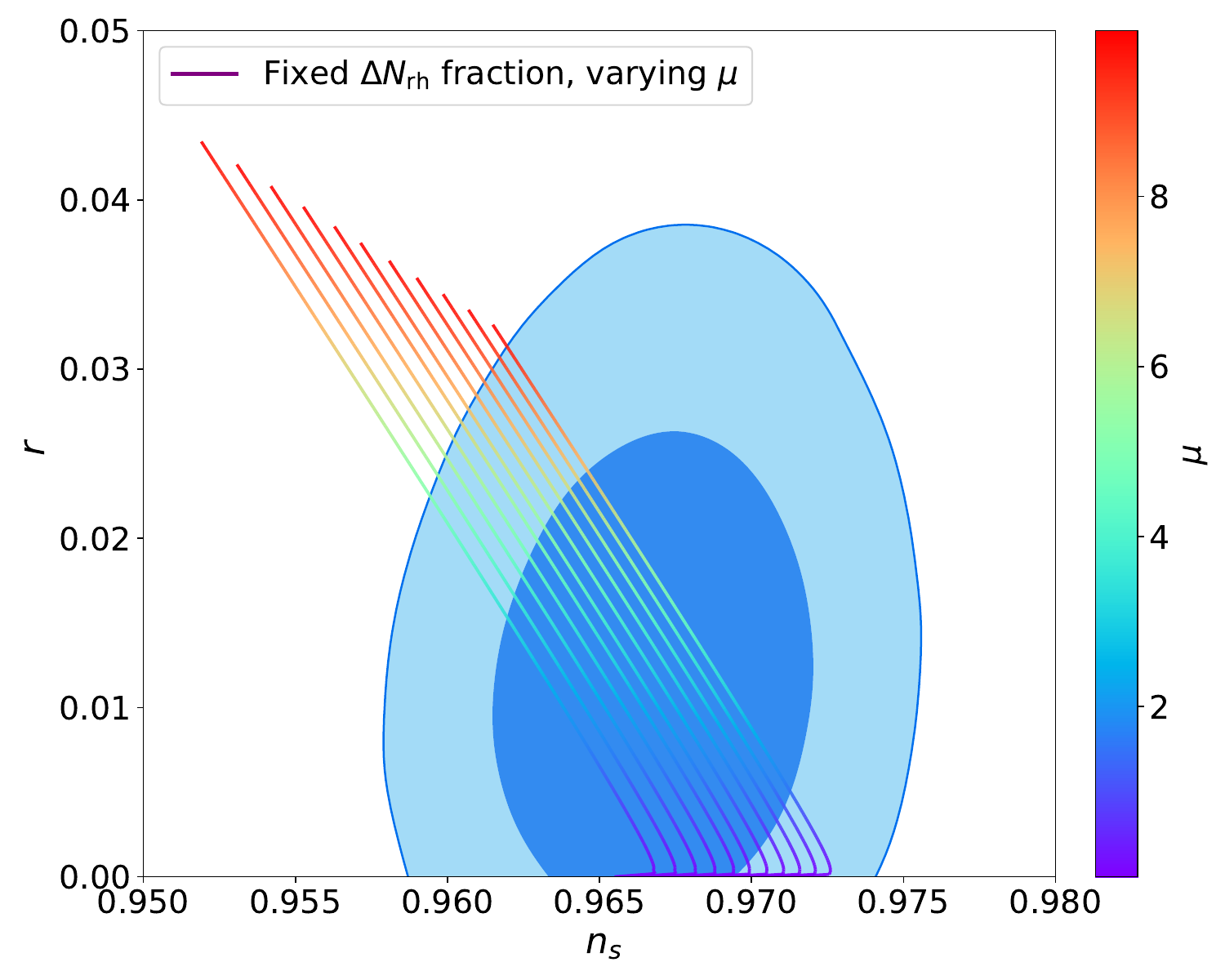}
    \end{subfigure}

    \caption{The predictions for \(n_s\) and \(r\) for the P-model of Eq. \eqref{eq:Polynomial-Attractor Model} with \(q=2\), for different values of \(\mu\). The reheating duration is varied within \(0\leq\Delta N_{\rm rh}\leq\Delta N_{\rm rh,max}(\alpha)\), and the predictions are calculated using the improved extended and universal expressions in Eq. \eqref{eq:re_poly_gen_ns} and Eq. \eqref{eq:re_poly_gen_r}. The $\Delta N_{rh}$ increases from right to left and moving along a curve of fixed $\Delta N_{rh}$ decreases the value of both the $\mu$ and the $\Delta N_{CMB}$. The light blue shaded region represents the 95\%  and the dark blue shaded region represents the 68\% confidence contours from Planck+BICEP/Keck 2018 data.}
    \label{non_universal_conture_plot}
\end{figure}

\begin{figure}[htbp]
    \centering

    \begin{subfigure}[b]{0.48\textwidth}
        \centering
        \includegraphics[width=\linewidth]{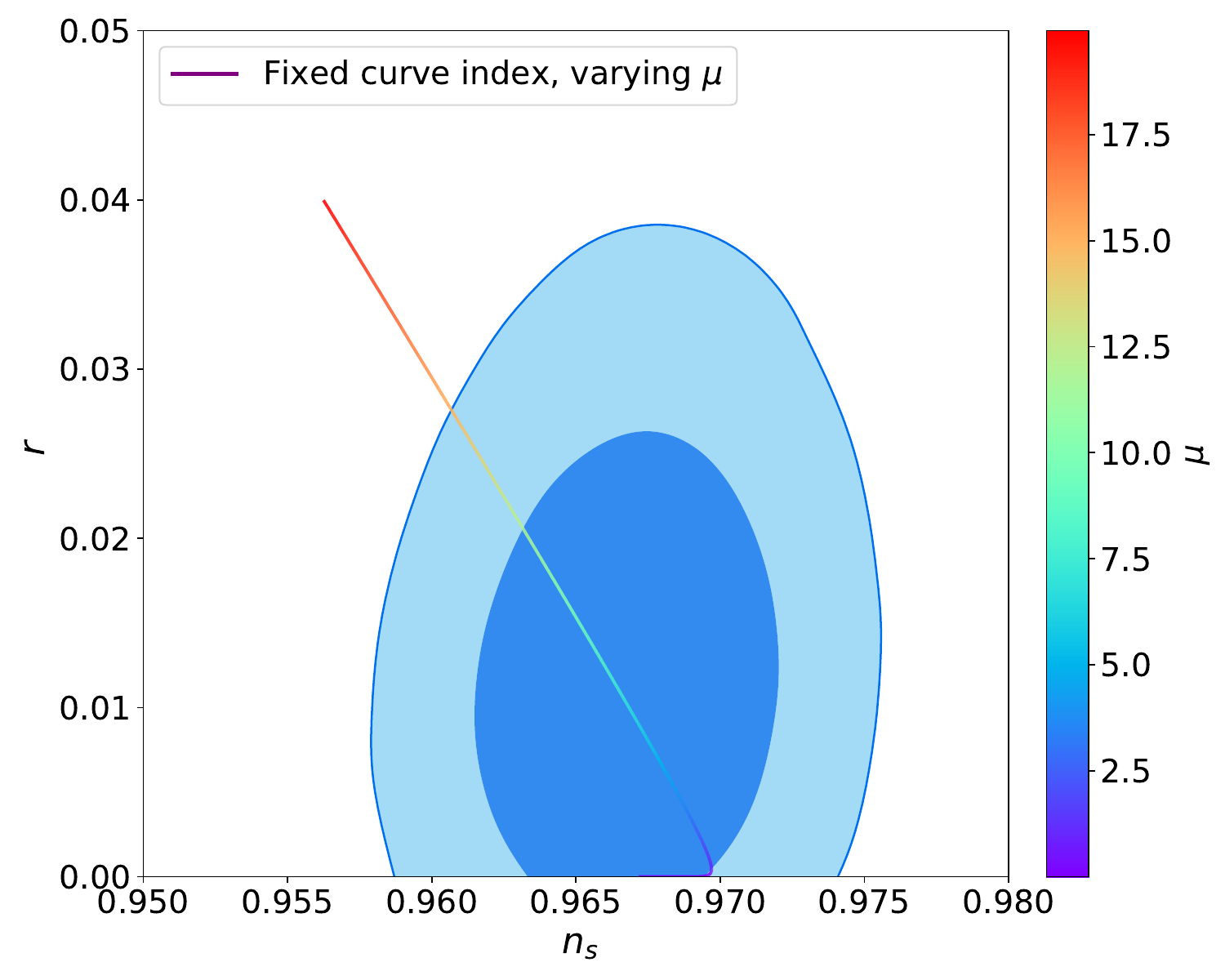}
    \end{subfigure}
    \hfill
    \begin{subfigure}[b]{0.48\textwidth}
        \centering
        \includegraphics[width=\linewidth]{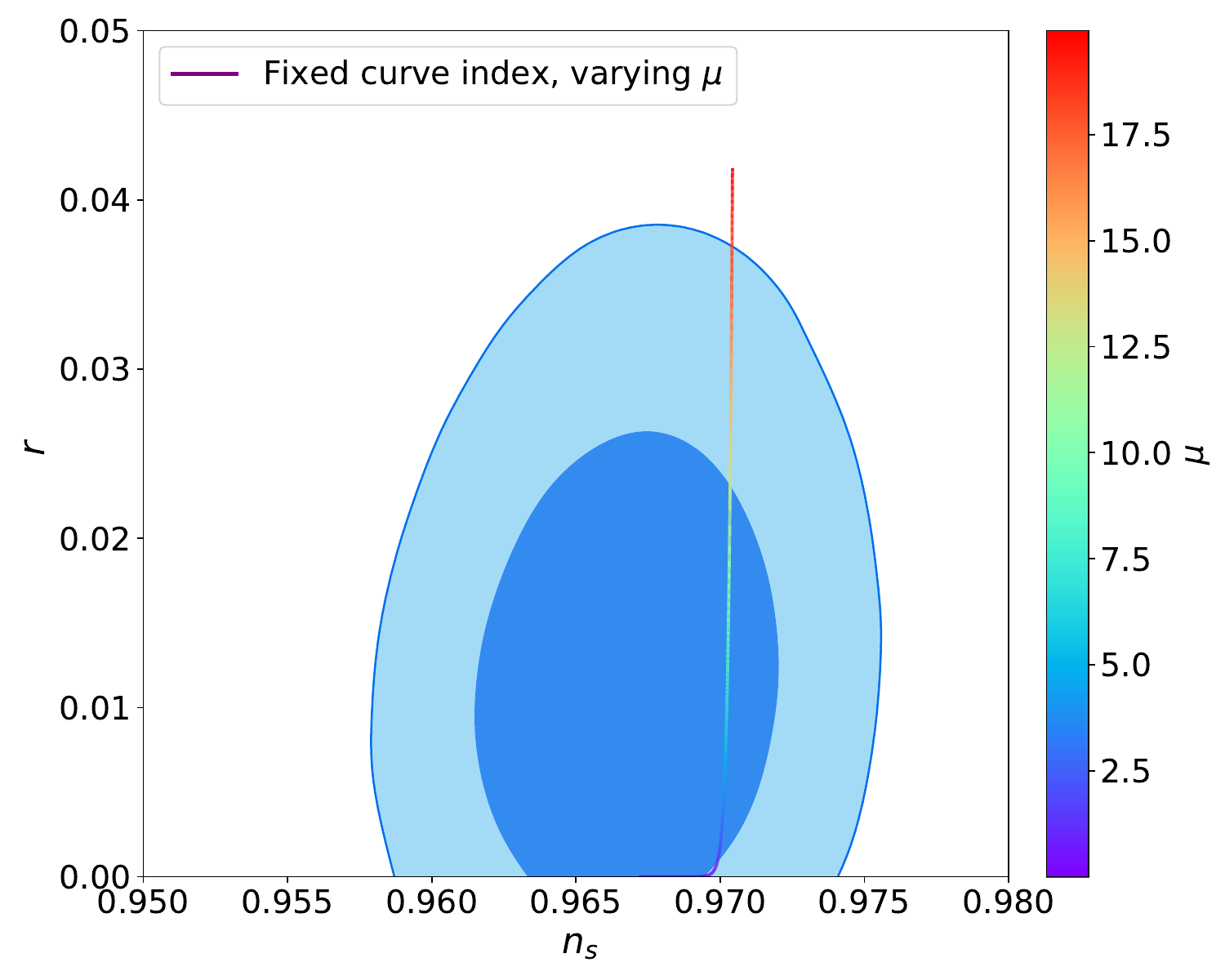}
    \end{subfigure}

    \caption{The predictions for the scalar spectral index \(n_s\) and tensor-to-scalar ratio \(r\) for the P-model given in Eq. \eqref{eq:Polynomial-Attractor Model} with \(q=4\) are shown for different values of the model parameter \(\mu\). The light blue shaded region represents the 95\%  and the dark blue shaded region represents the 68\% confidence contours from Planck+BICEP/Keck 2018 data.
}
    \label{q_4_universal_conture_plot}
\end{figure}

Mapping the predicted parameter space against the joint Planck+BICEP/Keck 2018 observational contours yields the results shown in Fig. \ref{non_universal_conture_plot} and \ref{q_4_universal_conture_plot}. For $q=2$, each curve corresponds to a specific value of $\Delta N_{\text{rh}}$, which increases from right to left, while $\mu$ and $\Delta N_{\text{CMB}}$ decreases from top to bottom along a fixed $\Delta N_{\text{rh}}$ line. In contrast, CMB data for $q=4$ allows the analysis of $\Delta N_{\text{CMB}}$ but not $\Delta N_{\text{rh}}$. Because $\Delta N_{\text{CMB}}$ does not depend on the reheating duration $\Delta N_{\text{rh}}$ in the $q=4$ case as shown in Fig.~\ref{q_4_universal_conture_plot}, the trajectories collapse into a single straight line parametrized by $\Delta N_{\text{CMB}}$ and $\mu$, rather than the family of curves observed for $q=2$.
\begin{table*}[htbp]
\centering
\caption{Model predictions for extended case for q=2 with different values of $\Delta N_{\rm CMB}$.
The observables $n_s$ and $r$, the model parameter $\mu$, and the reheating
e-folds $\Delta N_{\rm rh}$ are shown for
$\Delta N_{\rm CMB}=50$, $53$, and $55$.}
\label{tab:model_predictions}

\renewcommand{\arraystretch}{1.25}
\setlength{\tabcolsep}{8pt}
\setlength{\arrayrulewidth}{0.8pt}

\resizebox{\textwidth}{!}{%
\begin{tabular}{|cccc||cccc||cccc|}
\hline

\multicolumn{4}{|c||}{\textbf{$\Delta N_{\rm CMB}=50$}}
&
\multicolumn{4}{c||}{\textbf{$\Delta N_{\rm CMB}=53$}}
&
\multicolumn{4}{c|}{\textbf{$\Delta N_{\rm CMB}=55$}}
\\

\hline

$n_s$ & $r$ & $\mu$ & $\Delta N_{\rm rh}$
&
$n_s$ & $r$ & $\mu$ & $\Delta N_{\rm rh}$
&
$n_s$ & $r$ & $\mu$ & $\Delta N_{\rm rh}$
\\

\hline

0.9651 & 0.0137 & 3.47 & 23.456
&
0.9671 & 0.0128 & 3.55 & 11.480
&
0.9683 & 0.0123 & 3.59 & 3.4942
\\

0.9650 & 0.0138 & 2.51 & 22.458
&
0.9670 & 0.0130 & 3.59 & 11.481
&
0.9682 & 0.0124 & 3.63 & 3.494
\\

0.9620 & 0.0222 & 5.67 & 24.022
&
0.9644 & 0.0202 & 5.59 & 12.015
&
0.9659 & 0.0189 & 5.55 & 4.004
\\

0.9619 & 0.0224 & 5.71 & 24.037
&
0.9643 & 0.0204 & 5.67 & 12.016
&
0.9658 & 0.0190 & 5.59 & 4.0050
\\

\hline
\end{tabular}%
}
\end{table*}
\begin{table*}[htbp]
\centering
\caption{Model predictions for the universal case $q=2$ for different values of
$\Delta N_{\rm CMB}$. The scalar spectral index $n_s$, tensor-to-scalar ratio
$r$, model parameter $\mu$, and reheating e-folds $\Delta N_{\rm rh}$ are
shown for $\Delta N_{\rm CMB}=50$, $53$, and $55$. Note that no data points satisfying the 68\% confidence-level constraint on $\mu$ were found but have still been listed for completeness.}
\label{tab:q2_universal}

\renewcommand{\arraystretch}{1.25}
\setlength{\tabcolsep}{8pt}
\setlength{\arrayrulewidth}{0.8pt}

\resizebox{\textwidth}{!}{%
\begin{tabular}{|cccc||cccc||cccc|}
\hline

\multicolumn{4}{|c||}{\textbf{$\Delta N_{\rm CMB}=50$}}
&
\multicolumn{4}{c||}{\textbf{$\Delta N_{\rm CMB}=53$}}
&
\multicolumn{4}{c|}{\textbf{$\Delta N_{\rm CMB}=55$}}
\\

\hline

$n_s$ & $r$ & $\mu$ & $\Delta N_{\rm rh}$
&
$n_s$ & $r$ & $\mu$ & $\Delta N_{\rm rh}$
&
$n_s$ & $r$ & $\mu$ & $\Delta N_{\rm rh}$
\\

\hline

0.97000 & 0.00079 & 0.199 & 20.4890
&
0.971700 & 0.0010 & 0.279 & 8.005
&
0.972728 & 0.00089 & 0.1998 & 0.4278
\\

0.97002 & 0.0027 & 0.679 & 21.689
&
0.971701 & 0.0017 & 0.4795 & 9.3386
&
0.972731 & 0.00193 & 0.3594 & 1.476
\\

0.970005 & 0.0049 & 1.238 & 22.2912
&
0.971698 & 0.0029 & 0.79919 & 9.8723
&
0.972732 & 0.00318 & 0.9180 & 1.9707
\\

0.97001 & 0.0083 & 2.077 & 22.8754
&
0.971699 & 0.0048 & 1.3587 & 10.400
&
0.972727 & 0.0044 & 1.4385 & 2.4726
\\

\hline
\end{tabular}%
}
\end{table*}
\begin{table*}[htbp]
\centering
\caption{Model predictions for the extended case $q=4$ for different values of
$\Delta N_{\rm CMB}$. The scalar spectral index $n_s$, tensor-to-scalar ratio
$r$, and model parameter $\mu$ are shown for
$\Delta N_{\rm CMB}=54$, $55.5$, and $56$.}
\label{tab:q4_extended}

\renewcommand{\arraystretch}{1.25}
\setlength{\tabcolsep}{10pt}
\setlength{\arrayrulewidth}{0.4pt} 

\resizebox{\textwidth}{!}{%
\begin{tabular}{|ccc||ccc||ccc|}
\hline

\multicolumn{3}{|c||}{\textbf{$\Delta N_{\rm CMB}=54$}}
&
\multicolumn{3}{c||}{\textbf{$\Delta N_{\rm CMB}=55.5$}}
&
\multicolumn{3}{c|}{\textbf{$\Delta N_{\rm CMB}=56$}}
\\

\hline

$n_s$ & $r$ & $\mu$
&
$n_s$ & $r$ & $\mu$
&
$n_s$ & $r$ & $\mu$
\\

\hline

0.9692& 0.000011& 0.0399&
0.9694& 0.0016 &1.777&
0.9666&0.0105 &7.130 
\\
\hline
\end{tabular}%
}
\end{table*}
\begin{table*}[htbp]
\centering
\caption{Model predictions for the universal case $q=4$ for different values of
$\Delta N_{\rm CMB}$. The scalar spectral index $n_s$, tensor-to-scalar ratio
$r$, and model parameter $\mu$ are shown for
$\Delta N_{\rm CMB}=54$, $55.5$, and $56$.}
\label{tab:q4_universal}

\renewcommand{\arraystretch}{1.25}
\setlength{\tabcolsep}{10pt}
\setlength{\arrayrulewidth}{0.4pt}

\resizebox{\textwidth}{!}{%
\begin{tabular}{|ccc||ccc||ccc|}
\hline

\multicolumn{3}{|c||}{\textbf{$\Delta N_{\rm CMB}=54$}}
&
\multicolumn{3}{c||}{\textbf{$\Delta N_{\rm CMB}=55.5$}}
&
\multicolumn{3}{c|}{\textbf{$\Delta N_{\rm CMB}=56$}}
\\

\hline

$n_s$ & $r$ & $\mu$
&
$n_s$ & $r$ & $\mu$
&
$n_s$ & $r$ & $\mu$
\\

\hline

0.9692&0.0000112 &0.0399 &
0.9699&0.00148 &0.5978 &
0.9702&0.0107 &7.1302 
\\

\hline
\end{tabular}%
}
\end{table*}

At the pivot scale $k_{\text{CMB}} = 0.05\text{ Mpc}^{-1}$, Planck 2018 observations constrain the scalar spectral index to $n_s = 0.9649 \pm 0.0044$ ($68\%$ confidence), while combining with BICEP/Keck 2018 data sets an upper bound on the tensor-to-scalar ratio of $r_{0.05} < 0.036$ ($95\%$ confidence). Assuming the standard range $\Delta N_{\text{CMB}} \in [50, 60]$, the  allowed parameter ranges for $\mu$ and $\Delta N_{\text{rh}}$ were chosen accordingly instead of directly constrain $\mu$. For $q=2$, specific combinations of $n_s$ and $r$ following the BICEP/Keck 2018 likelihood for extended predictions are listed in Tab.~\ref{tab:model_predictions}. For the universal predictions, no data points satisfy the 68\% confidence-level constraint on $\mu$. However, several datasets that lie well within the 95\% confidence level are listed in Tab.~\ref{tab:q2_universal} for completeness. For $q=4$, the corresponding values of $n_s$, $r$, and $\mu$ across different values of $\Delta N_{\text{CMB}}$ are presented in Tab.~\ref{tab:q4_extended} for extended predictions and Tab.~\ref{tab:q4_universal} for universal predictions. For universal expressions of $n_s$ and $r$, the predicted curves as shown in Fig.~\ref{fig:univs nonuni} are nearly vertical, increasing $\mu$ raises $r$ while $n_s$ remains essentially fixed at the value set by the reheating fraction, meaning constraints on $\mu$ originate solely from the tensor-to-scalar ratio. Conversely, extended expressions yield tilted curves as shown in Fig.~\ref{non_universal_conture_plot} characterized by a roughly linear anti-correlation between $n_s$ and $r$, where higher values of $\mu$ increase $r$ while decreasing $n_s$. And as $\mu$ grows, the theoretical trajectories deviate rapidly from the preferred Planck+BICEP/Keck likelihood region.
 This behavior demonstrates that accounting for the full dependence of $\mu$ and $\Delta N_{\text{rh}}$ on $\Delta N_{\text{CMB}}$ imposes significantly tighter, joint constraints on $\mu$ and also leads to refined analytical predictions, which in turn improve the $n_s$ and $r$ values. Although a full Markov Chain Monte Carlo (MCMC) analysis was not performed in this study to directly constrain $\mu$, we evaluated the allowed parameter ranges. A dedicated MCMC sampling will be conducted in future work to establish precise statistical constraints on $\mu$.

\section*{Acknowledgement}
AS and AA acknowledge the computational facilities provided by IIT Mandi. AA acknowledges financial support from the University Grants Commission (UGC) under Fellowship Award No. 231610040032. The authors sincerely thank Rahul Kothari and Nitesh Kumar for valuable discussions and constructive suggestions, and Krishna Mohan Parattu for setting the initial research direction, identifying the problem, and providing helpful comments and editorial suggestions during the preparation of the manuscript.



\printbibliography


\end{document}